# Spectroscopically forbidden infra-red emission in Au-vertical graphene hybrid nanostructures

A. K. Sivadasan,[1,*,#] Santanu Parida,[1] Subrata Ghosh,[2,*] Ramanathaswamy Pandian,[1] and Sandip Dhara[1,*]

[1]*Nanomaterials Characterization and Sensors Section, Surface and Nanoscience Division, Indira Gandhi Centre for Atomic Research, Homi Bhabha National Institute, Kalpakkam-603102, India*

[2] *Thin Films Coatings Section, Surface and Nanoscience Division, Indira Gandhi Centre for Atomic Research, Kalpakkam-603102, India*

E-mail: sivankondazhy@gmail.com, subrataghosh.phys@gmail.com, dhara@igcar.gov.in

[#]Presently at Government Higher Secondary School, Pazhayannur -680587, Kerala, India

**Abstract**

Implementation of Au nanoparticles (NPs) is a subject for frontier plasmonic research due to its fascinating optical properties. Herein, the present study deals with plasmonic assisted emission properties of Au NPs-vertical graphene (VG) hybrid nanostructures. The influence of effective polarizability of Au NPs on the surface enhanced Raman scattering and luminescence properties is investigated. In addition, a remarkable infra-red (IR) emission in the hybrid nanostructures is observed and interpreted on the basis of intra-band transitions in Au NPs. The flake-like nanoporous VG structure is invoked for the generation of additional confined photons to impart additional momentum and a gradient of confined excitation energy towards initiating the intra-band transitions of Au NPs. Integrating Au plasmonic materials in three-dimensional VG nanostructures enhances the light-matter interactions. The present study provides a new adaptable plasmonic assisted pathway for optoelectronic and sensing applications.

## 1. Introduction

The localized surface plasmon resonance (LSPR) of plasmonic materials find widespread applications by either confining electromagnetic waves in nanosize regime and/or coupling of light with optoelectronic circuits [1-3]. Therefore, the plasmonic noble metallic nanostructures are considered as the bridge between photonics to electronics. Au nanoparticles (NPs) are one of the most promising noble metallic structures owing to its chemical inertness to oxygen and bio-organic molecules. In addition, it is easily attachable to the organic and biological molecules using simple physisorption or functionalization [4]. Au NPs have also the ability to confine light in visible regime and enhance electric field density at the vicinity of NPs. Therefore, Au NPs induced plasmonic effects are highly demanding in the field of spectroscopic investigations in terms of surface enhanced Raman spectroscopy (SERS) and tip enhanced Raman spectroscopy (TERS) [5,6]. The number density, size and distribution of Au NPs determine the polarizability and hence in-turn improve the enhancement factor of electromagnetic field. The dielectric constant of Au NPs and surrounding media also plays a significant role for enhancing the electromagnetic field density [1,5]. The improved light-matter interaction can be achieved by trapping the photons in the contrast of different dielectric materials. Recent reports related to the proximity induced light-matter interaction shed a glimpse of light to the Raman spectroscopic studies [7], and photoluminescence (PL) imaging of single semiconductor nanostructure (~100 nm) in the sub-diffraction limit [8]. In order to improve the optical gain, it is customary to design hybrid systems with the combination of optical confinement by means of dielectric contrast and plasmonics [9]. Even though, there are several studies available related to the electromagnetic enhancement properties [1-3], the investigations related to the emission properties of Au NPs, especially in the hybrid systems, are very limited. Therefore, the studies of luminescence properties of Au NPs over a broad regime, visible to infrared (IR), are highly demanding subject in the emerging area of plasmonic hybrid systems.

In the plasmonic hybrid systems, covalently bonded material such as organic and biological samples exhibiting significant change in polarizability, compared to that of ionic bonded ones, leads to improved light scattering efficiency [10]. In this context of higher change in polarizability, carbon materials are extensively used to experience the plasmonic effects easily. Moreover, carbon nanomaterials are very important in terms of their applications in electrochemical devices, sensors and field emission [11]. Among them, graphene is one of the promising two-dimensional (2D) materials because of the desired properties including high electrical and thermal conductivity, 97% transparency, and high mechanical stability [12]. However, an electronic band gap is essential for the emission properties of material towards its opto-electronic devices applications. A perfect single layer graphene does not show any PL properties because of the lack of electronic bandgap [11,12]. However, band gap could be created in graphene by several methods including cutting the edges of graphene [13], functionalizing the carbon nanostructures [14], creating defects or by fabricating multilayer graphene [11]. In this scenario, other carbon nanostructures, such as graphene nanoribbon, graphene oxide, flake-like vertical graphene (VG), carbon nanotube, carbon quantum dot, are the choice of material to fabricate hybrid structures.

Among them flake-like VG can be considered a potential candidate to utilize the effective plasmonic properties of Au NPs due to its unique orientation, high edge density, large surface area, presence of multilayers, ion–induced defects during growth, excellent optical, better thermal properties, capability of easy functionalization and chemical stability [15-18]. The presence of optical band gap may induce the emission properties in VG. Thus, VG itself a promising material for white light emission [19]. The emission properties can be enhanced through elevating the electronic transition probabilities by means of plasmonics. In addition, VG serves as an excellent mechanical backbone for other nanostructures [20,21]. The large surface area of VG is advantageous for the decoration of large number of Au NPs as a hybrid structure and contributes higher SERS as well as PL enhancement [22]. Further, size and inter-distance of Au NPs on VG found to have significant impact on the emission properties, as well [23]. The presence of Au NPs in VG nanoporous structures may not only confine the light but also can influence the PL emission properties of the entire hybrid system, which is not investigated in details. Thus, the graphene-plasmonic hybrid nanostructures can find various applications in enhanced light emitters, waveguide systems in nanosize regime, optical sensors in biomedical diagnosis, chemical- and bio-sensor, energy storage or conversion, food safety, photo-catalytic activities and environmental monitoring device [24-28].

The present study is mainly focused on the Raman and PL properties of Au NPs-nanoporous VG hybrid nanostructures. The role of effective polarization of Au NPs on the SERS and PL enhancement of VG is investigated. Importantly, continuum emissions from visible to IR regime of the Au NPs, which originate because of inter- and intra-band transitions, are identified. The flake-like structure in hybrid nanostructures is invoked to (i) explain the additional confinement of light photons for imparting the higher momentum to evanescent waves, (ii) the generation of a gradient of electric filed density and (iii) initiate the intra-band transitions by elevating the probability of transition rate to induce the IR emissions in Au NPs.

## 2. Experimental section

*2.1. Growth of VG by ECR-CVD*

The synthesis of VG was performed in electron cyclotron resonance plasma enhanced chemical vapor deposition (ECR-CVD). Carbon paper and $SiO_2$ are opted as substrate for the VG growth. The details of the growth are provided elsewhere [29]. In brief, the reaction chamber was evacuated down to a base pressure of $5 \times 10^{-6}$ mbar by a turbomolecular pump prior to growth. Subsequently, the substrates were pre-cleaned by exposing Ar plasma. Thereafter, $CH_4$ of 5 sccm along with Ar of 20 sccm was fed into the chamber for 60 min at a microwave power of 375 W. The substrate temperature of 800 °C and operating pressure of $1.2 \times 10^{-3}$ mbar were maintained during the growth for 30 min.

*2.2. Preparation of Au NPs-nanoporous VG hybrid system*

Commercially available high pure Au NPs colloidal solution was used for preparing the hybrid system with VG. The colloidal solution with two different concentrations was embedded in VG nanoporous structures to prepare the samples with different size and distribution of Au NPs. The uniform dispersion of Au NPs over the VG samples was achieved by drop casting followed by heating with an IR lamp (150W) for 10 min. The same procedures were followed for all samples used in the present study namely, VG/Carbon paper, VG/SiO$_2$ as well as planar nanographite (PNG)/SiO$_2$.

*2.3. Characterization techniques*

Morphological features of the VG, PNG, Au NPs-nanoporous VG and Au NPs-PNG were inspected using field emission scanning electron microscope (FESEM; Supra 55, Zeiss). The vibrational analyses as well as PL studies were carried out using a micro-Raman spectrometer (inVia Renishaw, UK) with excitation laser sources of wavelength 514.5 and 532 nm. The laser beam was focused and collected by using an objective lens of 100× magnification and a numerical aperture of 0.85. A grating with 1800 gr.mm$^{-1}$ was used for the monochromatization of scattered signal and a thermoelectrically cooled CCD detector was used for recording the spectra in the backscattered configuration.

## 3. Results and Discussions

*3.1. Morphological analysis*

The morphological analysis using FESEM images of as-prepared VG grown on carbon paper is shown in the figure 1. The low magnification image of the pristine sample, referred as $C_0$ (figure 1a; $C_0$), confirms the uniform coating of VG over the entire substrate. A magnified image, as depicted in figure 1b; $C_0$, indicates the flake-like structure of pristine VG.

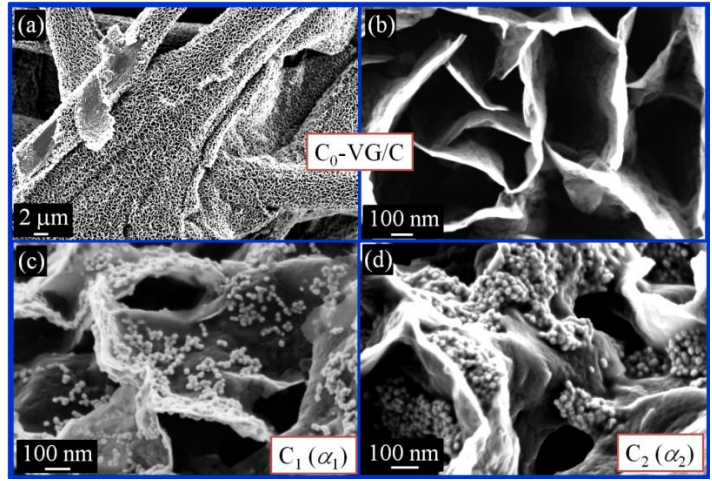

**Figure 1.** (a) FESEM images of flake-like nanoporous VG grown on carbon paper, (b) High magnification images of as-prepared nanoporous VG. High magnification images of VG decorated with Au NPs with polarizability per unit area of (c) $\alpha_1$ and (d) $\alpha_2$.

High magnification image is shown for (figure 1c) Au NPs-nanoporous VG hybrid nanostructures with low number density of Au NPs, labeled as $C_1$. Figure 1d displays the magnified image of Au NPs-VG hybrid nanostructures with comparatively higher number density of Au NPs, referred as $C_2$, throughout the manuscript.

### 3.2. Polarizability dependent Raman studies

The polarizability; $\alpha_{Au}$, a detrimental factor for the plasmonic interaction of Au NPs with the materials can be calculated using the following relation [1,5,30];

$$\alpha_{Au} = 4\pi a^3 \frac{\varepsilon_{Au} - \varepsilon_d}{\varepsilon_{Au} + 2\varepsilon_d} \quad\text{------(1)}$$

where, dielectric constant of Au ($\varepsilon_{Au}$) for the excitation laser of 514.5 nm is considered as ~ -3.541 and $\varepsilon_d$ is the dielectric constant of the surrounding media (air; $\varepsilon_d$=1). A statistical analysis was carried out for computing the number of Au NPs per area of 100×100 nm$^2$ from the micrograph (figure 1c and figure 1d) of hybrid nanostructures. In the case of sample $C_1$, number of particle per unit area ($n_1$) is obtained as ~10 (figure 1c) with the approximate radius of ~9 nm (supplementary material figure S1a). Using the equation 1, the polarizability for a single Au NP in $C_1$ ($\alpha_{Au}|_{a=9nm}$) is found as $\approx 27 \times 10^3$ nm$^3$. The corresponding effective polarizability for Au NPs per unit area of $C_1$ is; $\alpha_1 = n_1 \times \alpha_{Au}|_{a=9nm} \approx 27$ nm. In the case of $C_2$, number of particle ($n_2$) per area is obtained as ~33 (figure 1d) with an approximate radius of ~12.5 nm (supplementary material figure S1b). The polarizability for a single Au NP in the sample $C_2$ is calculated as $\alpha_{Au}|_{a=12.5nm} \approx 72 \times 10^3$ nm$^3$ and the corresponding effective polarizability ($\alpha_2$) is estimated as; $\alpha_2 = n_2 \times \alpha_{Au}|_{a=12.5\ nm} \approx 237.6$ nm. Therefore, the estimated polarizability per unit area of $C_2$ is about 9 times higher compared that of $C_1$ (*i.e.*, $\alpha_2 \sim 9\alpha_1$). To investigate the role of effective polarizability of Au NPs in hybrid system, vibrational properties of $C_0$, $C_1$ and $C_2$ are probed by the Raman spectra and depicted in figure 2a.

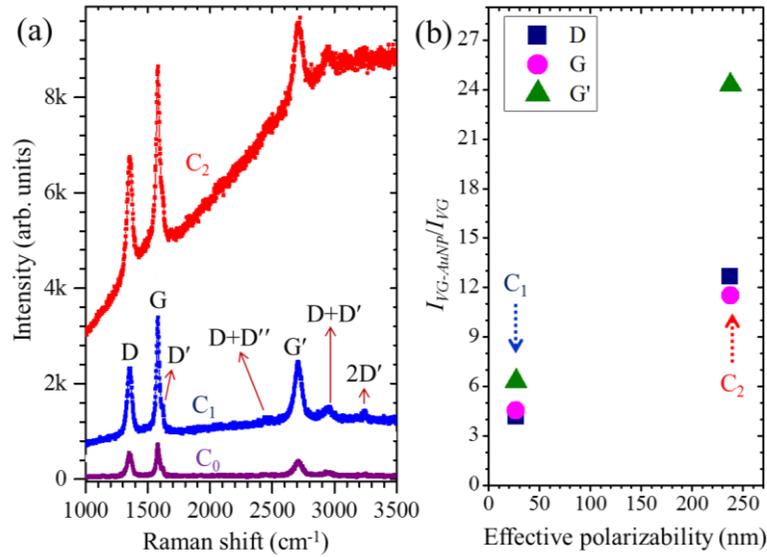

**Figure 2.** (a) Raman spectra of Au NPs-VG hybrid nanostructures. (b) Ratio of peak intensity of the plasmonic enhanced Raman modes to the normal Raman modes for the samples $C_1$ and $C_2$.

The typical Raman spectra consist of prominent intensities for D, G, and G′ bands and these phonon modes are considered as the main characteristic feature of VG [19,29]. The presence of G and G′ bands in the spectra (figure 2a) confirm the existence of graphitic structure. The phonon modes denoted as D, D′, D+D″, D+D′and 2D′ in the spectra (figure 2a) indicate the presence of different defects in VG [17]. These defect related modes originated due to large amount of edge states present in the flake-like VG structures, NG base layer with *sp*$^3$ bonded C–H species and ion-induced defects from plasma during the growth. In order to quantify the Au NP induced SERS effect, the intensity ratios of prominent Raman modes of the hybrid system ($I_{AuNP-VG}$) to that of pristine VG ($I_{VG}$) for the samples $C_1$ and $C_2$ are plotted in the figure 2b. It is clearly observable that the intensity ratios for different prominent modes (D, G and G′) were enhanced with a factor of 4-6 and 12-24 for $C_1$ and $C_2$, respectively (figure 2b). The difference in intensity enhancement factor for different Raman modes in each system is due to the slight variation in fluctuation of electric susceptibility with respect to the incident and scattered wave direction. In other words, it is an indication of different Raman allowed modes exhibiting different response to the change in polarizability [10]. The observed enhancement (figure 2) in spectral intensity is due to the enhancement of strong local electric fields in the vicinity of Au NP-VG hybrid interface because of the coupling of excitation light with the plasmon resonances of metallic nanostructures [30-32]. In addition, one order higher intensity enhancement factor of $C_2$ compared to that of $C_1$ is attributed to the higher effective polarizability of plasmonic Au NPs in $C_2$ (*i.e.*, $\alpha_2 \sim 9\alpha_1$).

*3.3. Photoluminescence studies*

In addition to the SERS enhancement in VG, a drastic background change in the spectra of $C_2$ is observed with respect to that of $C_1$. To understand the effective polarizability dependent luminescence behavior, we have extended the similar experiments by recording the PL spectra from Au NPs-VG (figure 3). The corresponding PL spectra collected for all the samples ($C_0$, $C_1$ and $C_2$) are shown in the figure 3a. Along with the luminescence peak the prominent phonon modes are also observed in the PL spectra. A tremendous enhancement in the intensities of emission bands for hybrid system, $C_1$ and $C_2$, compared to that of $C_0$ is observed. Moreover, the enhanced PL intensity of $C_2$ is exhibited by one order compared to that of $C_1$ in accordance with estimated polarizability per unit area. The higher plasmonic induced enhancement in $C_2$ is because of the elevation of electronic transition probability in presence of higher electric filed density. In order to check the possibility of substrate induced effects in the luminescence properties of hybrid system, similar experiments are repeated for the sample grown on $SiO_2$ substrate. The morphology of VG and Au NPs-VG with different effective polarizability is provided in the supplementary material (figure S2). Substrate independent morphology of VG and Au NPs-VG hybrid system is confirmed from the FESEM analysis (figure 1 and figure S2). Impressively, PL spectra (figure 3) of both sample grown on carbon paper (figure 3a) and $SiO_2$ (figure 3b) shows similar line shape and emission bands. Similarly, like previous samples grown on carbon paper, a large plasmonic assisted enhancement for the intensities of emission bands is observed (figure 3b) for $S_1$ and $S_2$ compared to that of $S_0$. Hence, substrate induced effects in the luminescence properties (figure 3) and phonon modes (figure 2 and supplementary material figure S3) are eliminated from our study.

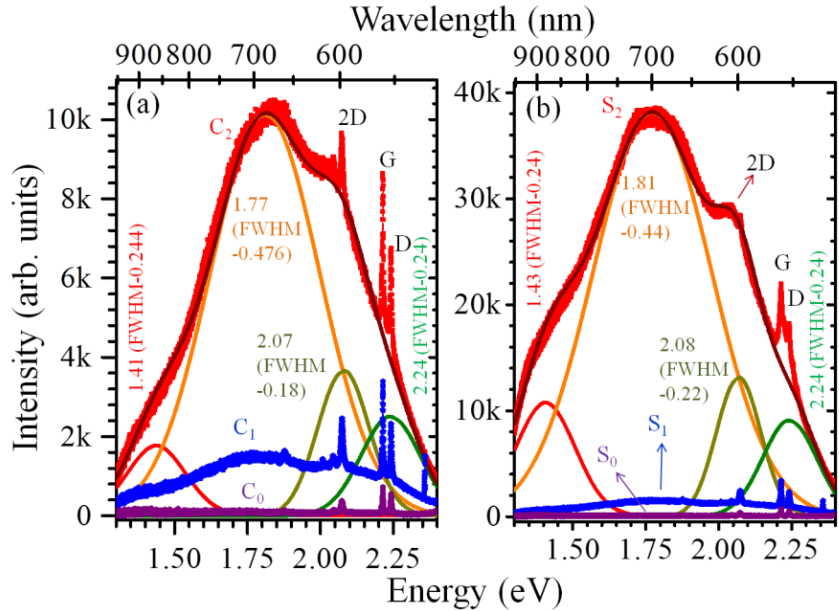

**Figure 3.** PL emission spectra of polarizability dependent plasmonic–VG hybrid nanostructures on the substrate of (a) carbon paper ($C_1$ and $C_2$) and (b) $SiO_2$ ($S_1$ and $S_2$). Gaussian fits are shown only for the intense peaks corresponding to sample $C_2$ and $S_2$.

The flake-like VG structure are expected to exhibit the PL peak at 1.89 eV (657 nm) and 2.24 eV (549 nm) within the energy range of 1.3 to 2.4 eV. However, PL emission of VG observed in present study is very week because of the thick layer of vertical sheets and presence of huge dangling bonds [33]. The PL peaks of VG are clearly visible in the hybrid system due to the plasmonic effects of Au NPs. The observed emission bands for the $C_2$ sample is deconvoluted using Gaussian function (figure 3). The major emission bands observed at around 1.41, 1.77, 2.07 and 2.24 eV with a full width at half maximum (FWHM) of approximately 0.24, 0.45, 0.20 and 0.24 eV, respectively. The PL peak centered at ~2.24 eV originated due to the transition between the $\pi^*$ and $\pi$ bands of VG, or the transition between the $\pi^*$ band and valence band formed by defects in VG. The PL peak at 1.77 eV originated due to the transition between the $\pi^*$ band and valence band formed by the oxygen defects in VG [19,35]. The presence of oxygen is obvious, since atmospheric moisture can be absorbed on the exposed surface edges of VG [35]. Other prominent PL peaks at 2.07 and 1.41 eV may be generated from the Au NPs. To ensure it, PL spectra from Au NPs decorated on $SiO_2$ substrate is compared with those for Au NPs-VG hybrid nanostructures on carbon ($C_2$) and $SiO_2$ substrates ($S_2$) with an effective polarizability of $\alpha_2$ (figure 4a). In addition to that, PL spectra are recorded directly from the Au NPs colloidal solution (figure 4b). The Gaussian fitted PL emission band from the Au NPs colloidal solution (figure 4b) exhibited continuum emission band centered around 2.01 eV with a FWHM of ~0.51 eV. Hence, the band at 2.01 eV for the Au NPs-nanoporous VG hybrid nanostructures is attributed to the broad visible continuum emission from the Au NPs (Figs. 3 and 4a). However, no noticeable PL emission at 1.43 eV is observed for Au NPs alone (figure 4b). Noteworthy to mention that, a strong IR emission centered around 1.43 eV is only observed in Au NP– nanoporous VG hybrid system (Figs. 3 and 4a). A strong plasmonic interaction with a significant shift in the LSPR peak is observed for the Au NPs-VG in absorption studies (supplementary material figure S4). Considering large band width and shift of the LSPR peak, typical PL emission for the hybrid nanostructure is also acquired using 532 nm excitation to show the presence of IR emission in Au NP– nanoporous VG hybrid system (supplementary material figure S5).

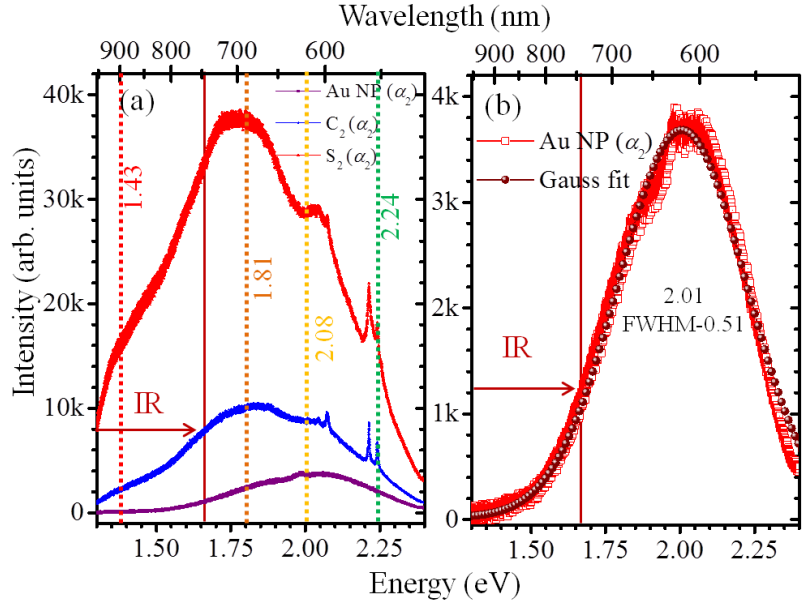

**Figure 4.** (a) Comparative plot of PL emissions of Au NPs on SiO$_2$, Au NPs-VG on carbon paper (C$_2$) and on SiO$_2$ (S$_2$) with an effective polarizability of $\alpha_2$ (b) PL emission spectra from direct Au NPs colloidal solution.

In order to investigate the influence of morphology of VG on the PL emission bands observed in the IR regime, similar experiments were carried out for Au NPs-PNG hybrid structure. Figure 5a depicts the high magnification FESEM images of PNG grown on the SiO$_2$ substrate. The corresponding Raman spectrum (figure 5b) of PNG is very similar to that of VG. Here, high intense D-band and other defect related peak originated due to the boundary-like defects and ion-induced defects by plasma during the growth [35,36]. The panel (c) in figure 5 shows the morphology of PNG decorated with Au NPs. The comparison for the PL spectra of Au NPs-PNG and Au NPs-VG hybrid nanostructures is shown in the figure 5d. The broad and continuum emission in the visible regime are observed for the both of the Au NPs-PNG and Au NPs-VG system (figure 5d). The IR emission band at 1.43 eV is observed only for the sample possesses the flake-like VG structure (figure 5d).

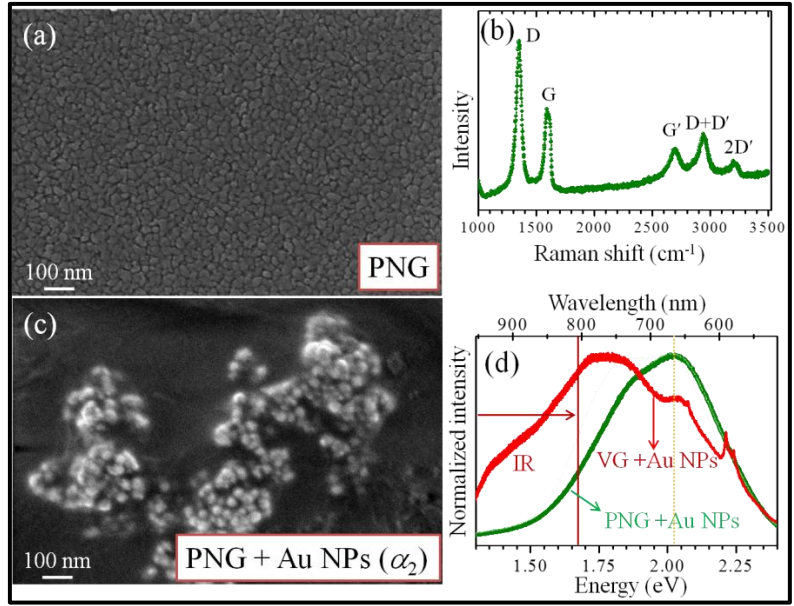

**Figure 5.** (a) FESEM images of PNG grown on SiO$_2$ substrate, (b) Raman spectrum of PNG, (c) FESEM image of Au NPs decorated PNG, (d) Comparative PL spectra of Au NPs-PNG and Au NPs-VG hybrid nanostructures.

Looking back to PL spectra of Au NPs-VG (figure 3), the prominent peak at 1.43 eV observed only in hybrid system may be originated from intra-band transition of Au NPs. The model reported for the intra band transition in Au NPs proposes that the low energy emission (0.81 eV) is originated from radiative transitions within the *sp*-conduction band across highest occupied molecular orbital (HOMO) and lowest unoccupied molecular orbital (LUMO) energy levels [37,38]. The broad visible continuum emission in Au NPs is already well established. The origin of this continuum emission is due to the inter-band transitions of electrons from *d* band to an unoccupied state of partially filled *sp*-conduction band and followed by radiative recombination process (supplementary material figure S6) [30,39].

In band diagram of Au [39], the responsible states for intra-band transitions are significantly separated in the momentum space (supplementary material figure S6). Generally, the wave vectors of visible light participated in inter-band transitions is comparatively negligible with respect to that of an electron. The radiative recombination process of electrons and holes or excitons demands the momentum conservation; $\vec{k}_e - \vec{k}_h = \vec{k}_{photon} \approx 0$, where, $\vec{k}_e, \vec{k}_h$ and $\vec{k}_{photon}$ are the wave vectors of electrons, holes and photons, respectively. Hence, the photon momentum can be neglected as far as the light-matter interaction, involved in the radiative recombination process, is concerned [40-42]. Therefore, the vertical transitions (supplementary material figure S6) are only allowed in the case of inter-band transition and it cannot be accounted for the IR emission of Au NPs [30,37-40]. Whereas, for IR emission, the radiative recombination may take place due to intra-band transitions between the two states located in *sp*-conduction band itself (supplementary material figure S6). However, the emission spectra of Au NP-VG hybrid system are strongly influenced by LSPR and it is obvious that the IR emission is not a red tail of the visible PL spectrum of Au (figures 3 and 4a). Therefore, we concluded that the strong IR emission is not because of the inter-band transitions, but it is generated due to the intra-band transitions. To the best of our knowledge, the spectroscopically forbidden IR emission of Au NP in hybrid system is experimentally demonstrated for the first time. In general, the intra-band transitions are spectroscopically hidden because of the two major reasons. Firstly, the direct intra-band optical transition is dipole forbidden because of the initial and final electronic states are located in same *sp*-conduction band, and therefore it possesses same symmetry. Secondly, the transition states are significantly separated energy states in momentum space, thus a normal photon momentum cannot provide the corresponding wave vector difference to preserve momentum conservation rule [39,40]. Therefore, it is inferred that the PL emission at IR regime (1.43 eV) can be enhanced by effectively decorating Au onto the suitable platform like nanoporous VG.

The scattering of excitation light from nanostructured surfaces, sub-wavelength protrusions and nanosized holes provides additional momenta for evanescent waves to generate LSPR in the Au NPs or its hybrid structures [43]. However, the evanescent fields near to metal nanostructures possess wave numbers that can be large enough to span momentum space between the initial and final states of a near-IR transition. Furthermore, the interaction of excitation with Au NPs leads to a very strong confinement of electromagnetic field due to LSPR. The order of spatial confinement light is equivalent to its physical dimensions. The folding of graphene layer or formation of wrinkles, V-shaped edges, grooves and porous structures, flake-like structures, wedge geometries of graphene can able to confine the light too [44].

Consequently, the photons in the evanescent field at the vicinity of Au NPs-VG hybrid nanostructures gain a significant amount of additional momentum [3,4,9,44,45], such that it leads to the relaxation of momentum conservation rules for the intra-band electronic transitions in the Au NPs. In addition, the evanescent waves confined on the interface of the Au NP-VG hybrid system increase the multipole or higher order transitional moments, whose symmetry rules no longer prohibit intra-band transitions [39]. Therefore, the strong field confinement near Au NPs-VG hybrid nanostructures are accompanied by strong field gradients, implying that electric dipole selection rule may no longer be valid. Eventually, it can be concluded that the additional enhancement of intra-band transitions in Au NPs is because of the presence of localized field. The edges of flake-like VG with porous nature influence both the localized field intensity and its gradient. Therefore, the observation of IR emission from the plasmonic-hybrid nanostructures may be owing to the enhanced localization and confinement of electromagnetic field density, which can be a useful for the applications of light confining techniques.

## 4. Conclusions

The effect of polarizability of Au nanoparticles (NPs) on surface enhanced Raman scattering (SERS) for the phonon modes of plasmonic-vertical graphene (VG) hybrid system is investigated. The intensity ratio of prominent phonon modes in the VG is enhanced by one order in harmony with the polarizability per unit area. In addition to the polarizability dependent plasmonic assisted enhancement for the photoluminescence (PL) intensities, plasmonic hybrid system also exhibits continuum emission from IR to visible regime. The visible emission in hybrid system is originated due to the inter-band transitions, whereas IR emission is attributed to the intra-band transitions of Au NPs. Absence of IR emission in Au NPs-planar nanographite system emphasize the impact of flake-like VG structure. The flake-like structure of nanoporous VG provides an additional confinement of the light. The confinement helps the relaxation of constraints due to the momentum conservation rule by generating sufficient amount of evanescent photons. In addition, the strong electric field gradients near the Au NPs-VG leads to the generation of multipole transitions which no longer prohibits the electronic dipole selection rule. Therefore, we conclude that spectroscopically forbidden IR emission from the plasmonic-hybrid nanostructures is because of the presence of evanescent waves with higher momenta. The present scientific inference paves the way of plasmonic-hybrid nanostructures towards nanoscale light confinement based applications.

## 5. Acknowledgements

We thank Kishore. K. Madupu, NCSS, SND, IGCAR for his valuable suggestions and fruitful discussions.

# Supplementary Material

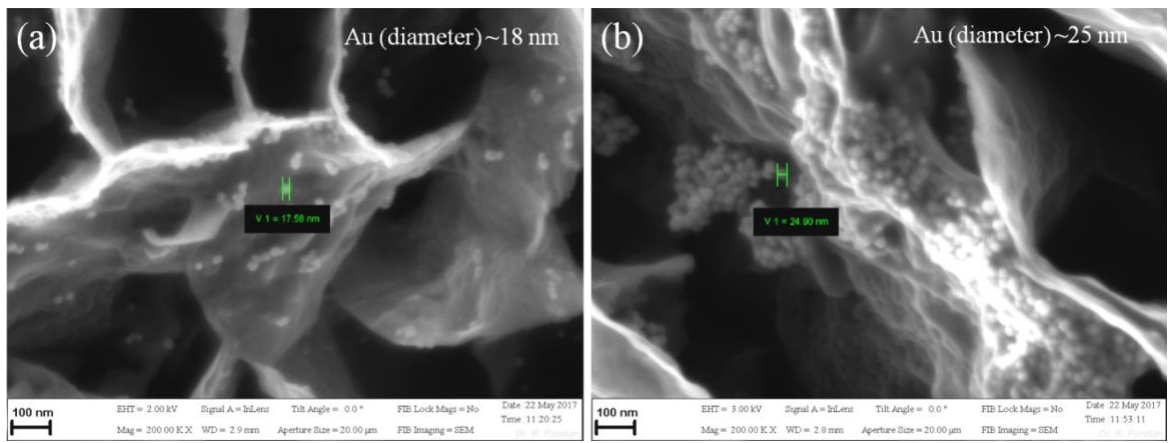

**Figure S1.** FESEM images of flake-like nanoporous VG on carbon paper. High magnification images of Au NPs with a diameter of (a) ~18 nm ; $\alpha_1$ and (b) ~ 25 nm; $\alpha_2$.

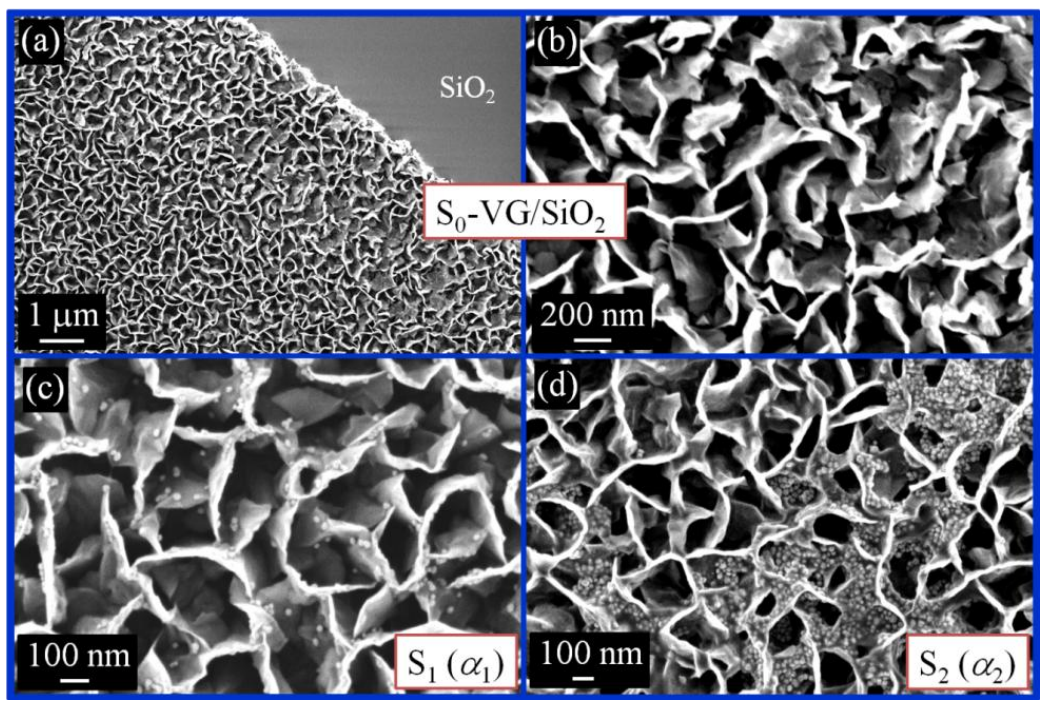

**Figure S2.** (a) FESEM images of flake-like nanoporous VG on $SiO_2$ substrate, (b) High magnification images of the as prepared nanoporous VG. High magnification images of Au NPs with an effective polarizability per area of 100 × 100 $nm^2$ of (c) $\alpha_1$ and (d) $\alpha_2$.

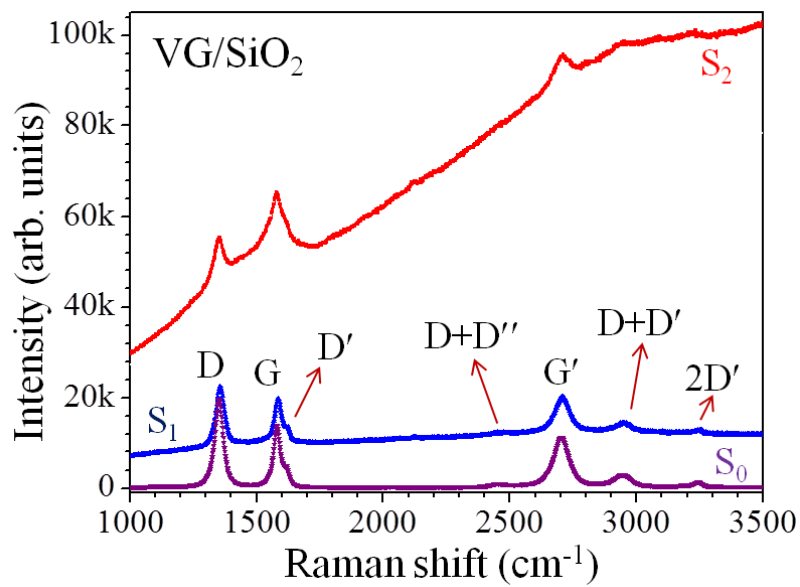

**Figure S3.** Raman spectra of (a) VG; $S_0$, (b) Au NPs-VG/SiO$_2$ with effective polarizability of $\alpha_1$ ; $S_1$ and (c) Au NPs-VG/SiO$_2$ with effective polarizability of $\alpha_1$; $S_2$.

Typical absorption spectra for 18 and 25 nm Au nanoparticles (NPs) on SiO$_2$, VG and Au NPs-VG hybrid structure are shown in the figure S4. In the spectrum of the Au NPs (figures S4a, S4b), peaks centered ~537 and ~540 nm with FWHM of ~70-75 nm is observed because of the localized surface plasmon resonance (LSPR) of the Au NPs of diameters 18 and 25 nm, respectively. The absorption spectrum of VG (figures S4c, S4d) shows broad peak centered ~700 nm. Whereas, in case of the Au NPs-VG hybrid structure the peaks correspond to the LSPR of Au NPs red shifts to ~565 and 590 nm for the Au NPs of diameters 18 and 25 nm, respectively (figures S4e, S4f). Thus, there is a strong plasmonic interaction of Au NPs and nonporous VG in the Au NPs-VG hybrid structure.

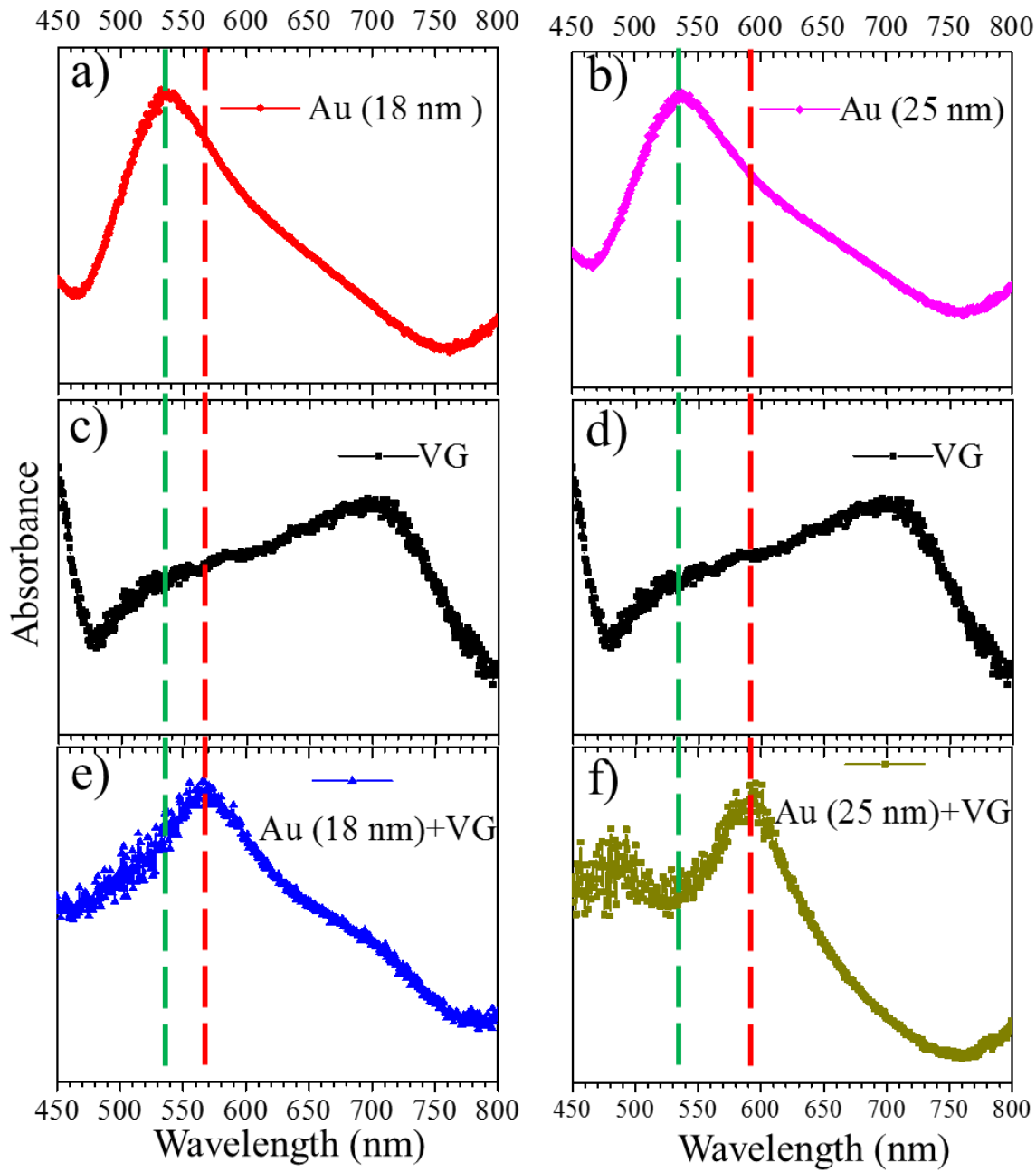

**Figure S4:** Absorption spectra of (a, b) Au NPs (c, d) VG and (e, f) Au NPs-VG hybrid structures recorded using an UV-Vis absorption spectrometer (Avantes) in the reflection geometry for the range of 450 to 800 nm. vertical dashed lines are guide to eye showing the red shift of LSPR peaks of hybrid Au NPs-VG structures with respect to that for Au NPs.

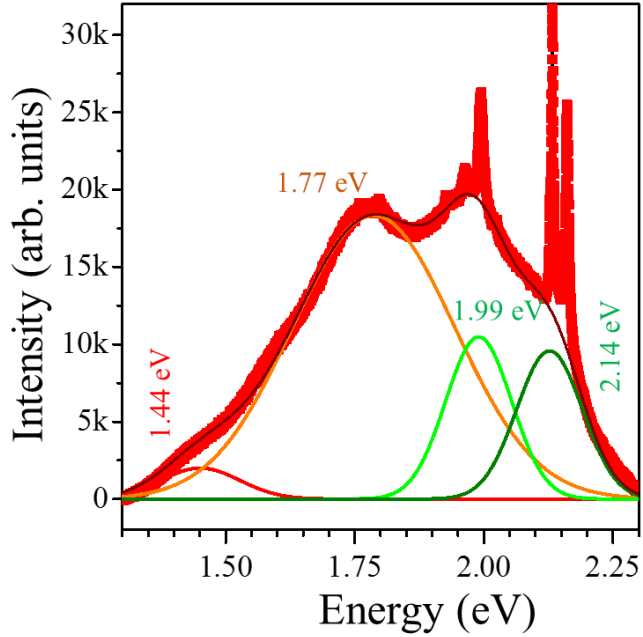

**Figure S5:** PL spectrum of Au -VG hybrid structure using 532 nm excitation. Sharp peaks correspond to the Raman modes of graphitic carbon with their usual nomenclature. The peaks ~1.41 and 1.99 eV correspond to the IR emission and inter-band transition, respectively from Au NPs. The PL peak ~1.77 eV originates due to the transition between the $\pi^*$ band and valence band formed by the oxygen defects in VG. The PL peak centered ~2.14 eV originates due to the transition between the $\pi^*$ and $\pi$ bands of VG, or the transition between the $\pi^*$ band and valence band formed by defects in VG.

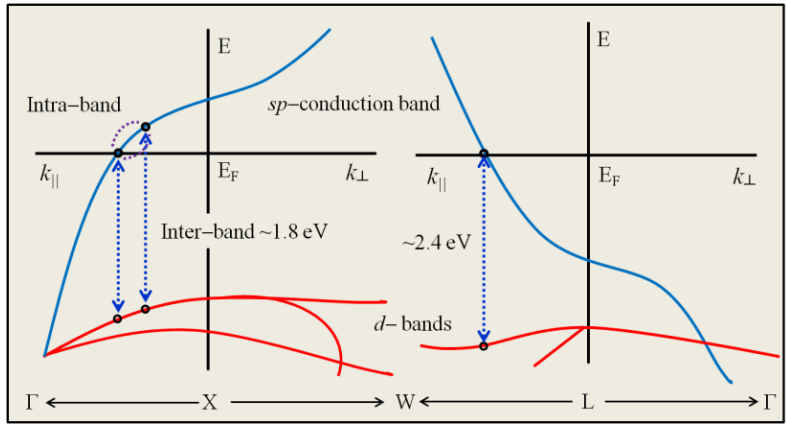

**Figure S6.** A schematic band diagram of Au depicting the inter- and intra- band transitions [1]. Copyright (2003) by the American Physical Society.

References:

[1] Beversluis M R, Bouhelier A and Novotny L 2003 Continuum generation from single gold nanostructures through near-field mediated intraband transitions *Phys. Rev. B* **68** 115433